\documentclass[fleqn,usenatbib]{mnras}

\usepackage{newtxtext,newtxmath}
\usepackage[T1]{fontenc}
\usepackage{ae,aecompl}
\usepackage{graphicx}	
\usepackage{amsmath}	
\usepackage{xcolor}
\usepackage{tablefootnote}

\newcommand{\sig}{\,$\text{cm}^{2}\, \text{g}^{-1}$}	

\title[SIDM haloes On FIRE]{The central densities of Milky Way-mass galaxies in cold and self-interacting dark matter models}

\author[Sameie et al.]{
Omid Sameie,$^{1}$\thanks{E-mail: sameie@utexas.edu}
Michael Boylan-Kolchin,$^{1}$
Robyn Sanderson,$^{2,3}$
Drona Vargya$^{2}$ 
\newauthor
Philip F. Hopkins,$^{4}$
Andrew Wetzel,$^{5}$
James Bullock,$^{6}$
Andrew Graus,$^{1}$
Victor H. Robles$^{7}$
\\
$^{1}$Department of Astronomy, The University of Texas Austin, 2515 Speedway, Stop C1400, Austin, TX 78712 USA\\
$^{2}$Department of Physics and Astronomy, University of Pennsylvania, 209 South 33rd Street, Philadelphia, PA 19104, USA\\
$^{3}$Center for Computational Astrophysics, Flatiron Institute, 162 Fifth Avenue, New York, NY 10010 USA\\
$^{4}$TAPIR, California Institute of Technology, Pasadena, CA 95616, USA\\
$^{5}$ Department of Physics and Astronomy, University of California, Davis, CA 95616, USA\\
$^{6}$ Department of Physics and Astronomy, University of California, Irvine, CA 92697, USA\\
$^{7}$ Yale Center for Astronomy and Astrophysics, New Haven, CT 06520, USA
}

\date{Accepted XXX. Received YYY; in original form ZZZ}

\pubyear{2020}

\begin{document}
\label{firstpage}
\pagerange{\pageref{firstpage}--\pageref{lastpage}}
\maketitle

\begin{abstract}
We present a suite of baryonic cosmological zoom-in simulations of self-interacting dark matter (SIDM) haloes within the ``Feedback In Realistic Environment'' (FIRE) project. The three simulated haloes have virial masses of $\sim 10^{12}\, \text{M}_\odot$ at $z=0$, and we study velocity-independent self-interaction cross sections of 1 and 10 ${\rm cm^2 \, g^{-1}}$. We study star formation rates and the shape of dark matter density profiles of the parent haloes in both cold dark matter (CDM) and SIDM models. Galaxies formed in the SIDM haloes have higher star formation rates at $z\leq1$, resulting in more massive galaxies compared to the CDM simulations. While both CDM and SIDM simulations show diverse shape of the dark matter density profiles, the SIDM haloes can reach higher and more steep central densities within few kpcs compared to the CDM haloes. We identify a correlation between the build-up of the stars within the half-mass radii of the galaxies and the growth in the central dark matter densities. The thermalization process in the SIDM haloes is enhanced in the presence of a dense stellar component. Hence, SIDM haloes with highly concentrated baryonic profiles are predicted to have higher central dark matter densities than the CDM haloes. 
Overall, the SIDM haloes are more responsive to the presence of a massive baryonic distribution than their CDM counterparts. 
\end{abstract}

\begin{keywords}
methods: numerical-galaxies: evolution-galaxies: formation-galaxies: structure- dark matter
\end{keywords}


\thanks{}
\section{Introduction}
The leading theory of structure formation assumes dark matter (DM) is cold and collisionless \citep{peebles1982,bond1982,blumental1984,davis1985}. Predictions of the cold dark matter (CDM) model are well tested on large scales \citep{seljak2006,percival2007,vog2014a,vog2014b,planck2018}, yet there are strong hints on the galactic scales which suggest the CDM model might fail to reproduce a handful of observations \citep[see][for a comprehensive review]{bullock2017}. These small-scale challenges include the CDM haloes from numerical simulations that are more massive than the observed population of galaxies around the Milky Way (MW) and in the Local Group  \citep{mbk2011,mbk2012,gk2014}, more flattened DM density profiles in the observed galaxies (core) vs. steeper profiles (cusp) from CDM-only simulations \citep{moore1994,kuzio2008,walker2011,oh2015}, and the diversity in the shape of the observed galactic rotation curves \citep{oman2015}. 

There are two general approaches to address the shortcomings of CDM-only simulations to accommodate these observations. Gravitational potential fluctuations caused by stellar feedback, either in the form of single feedback events (e.g., \citealt{navarro1996b}) or from repeated cycles of gas inflow and outflow (e.g., \citealt{mashchenko2008,pontzen2012,governato2012}), can account for the mass deficit in the observed dwarf galaxies. The critical requirements for the feedback to create extended DM cores are the total amount of energy injected by SNe events \citep{penarrubia2012}, bursty and extended star formation histories \citep{benitez2019}, and high density threshold for star formation $n_{\rm sf}\sim 10-1000\, {\rm cm^{-3}}$ \citep{hopkins2014,hopkins2018,bose2019} in galaxies that form their stars relatively late in their cosmological evolution, $z\leq 2$ \citep{onorbe2015}. Numerical simulations of galaxy formation indicate that feedback-induced core formation is most efficient at  $M_*/M_{\rm h}\sim 10^{-2}$; galaxies with lower baryon conversion efficiencies retain cuspier profiles, while more complicated effects occur at higher baryon conversion efficiencies. \citep{dicintio2014,chan2015,tollet2016,fitts2017,lazar2020}. 

In the second approach, basic assumptions about the nature of DM are revisited. For example, two-body non-gravitational interactions of DM particles have been introduced as a potential solution to the small-scale issues \citep{spergel2000,ahn2005,ackerman2009,arkani2009,feng2009, loeb2011,tulin2013}. N-body simulations of self-interacting dark matter (SIDM) models have shown that self-interactions would create constant density cores with spherical halo shapes in the core region \citep{vogel2012,zavala2013,rocha2013,peter2013}. Core formation in SIDM models is prompted by redistribution of energy and momentum of DM particles: dark matter scattering leads to a cored central density and an isothermal central temperature profile. As a result, cored DM densities are prevalent in DM-dominated SIDM systems \citep[see][ for a complete review]{tulin2018}.

Recent simulations of SIDM haloes in the presence of central baryonic components reveal that baryons modify this appealingly simple picture: once baryons dominate the central gravitational potential, a DM core will transform into a more cuspy profile  \citep{elbert2018,sameie2018,robles2019,despali2019}. The stellar concentration, along with the DM cross section, therefore plays a critical role in establishing the density profile of SIDM haloes  \citep{elbert2018,sameie2018}. An important result from these simulations is that SIDM can lead to a large diversity in the shape of rotation curves \citep{creasey2017,kamada2017, ren2019}, similar to what is observed in galaxy rotation curves \citep{lelli2016}.
The observed spread in the central DM densities of Milky Way (MW) satellites \citep{read2019,kaplinghat2019b} could also be triggered by SIDM  \citep{koda2011,zavala2019, kahlhoefer2019,sameie2020a,sameie2020b,correa2020}: core collapse in  tidally-evolving SIDM subhaloes can occur in much shorter time scales than in isolated SIDM haloes owing to mass loss in their outskirts. Nonetheless, most SIDM simulations have relied on some idealized realization of isolated galaxies or MW satellites in which either the effects of star formation and feedback are ignored or the parent haloes are treated as fixed potentials without considering their mass assembly and cosmological evolution.

Despite intriguing results, there have been comparably few attempts to run baryonic cosmological simulations in the context of SIDM models. There are simulations in the mass scales relevant to dwarf galaxies \citep{vog2014c,fry2015,fitts2019}, galaxies with halo masses few times $10^{12}\, \text{to}\, 10^{13} \text{M}_\odot$, representative of MW-mass and elliptical galaxies \citep{dicintio2017,despali2019}, galaxy clusters \citep{robertson2018,robertson2019}, and large-box cosmological simulations with spatially-uniform resolution \citep{lovell2018}. Most of these works either have medium-range mass resolutions or simulate their haloes with only one SIDM cross section. Thus far, there has not been any attempt at a suite of fully baryonic cosmological SIDM simulations with multiple cross sections and high mass resolution devoted to MW-mass galaxies and their satellites.

This paper studies the interplay of baryonic physics and DM self-interactions at the mass scales relevant to the MW. By employing the state-of-the-art FIRE-2 implementation of star formation and feedback, our simulations benefit from an explicit treatment of the interstellar medium (ISM) that resolves its multi-phase structure. Results from CDM simulations with this implementation agree with a broad range of observations including galaxy morphologies, the internal structure of the ISM, star formation histories, and the observed mass-size relation \citep{wetzel2016,hopkins2018b,hopkins2018}. Our simulation suite also has the high mass and spatial resolution ($m_{\rm dm}\sim 10^4\, \text{M}_\odot$, $\epsilon^{\rm min}_{\rm p}\sim 20\, {\rm pc}$) required to resolve the inner structure of the parent haloes. 

In this work, we focus on the stellar mass assembly and DM density profiles in our SIDM simulations and comparisons with their CDM counterparts; a companion paper (Vargya et~al., in preparation) employs the same simulation suite to study the shapes of dark matter haloes in SIDM and CDM, both with and without baryonic physics. We analyze in detail the cosmological evolution of both DM and stellar density profiles, and highlight the importance of concentration of stellar distribution on the diversity in DM density profiles. Our paper is organized as follows: in Section \ref{sec:sim} we give a brief description of our simulations. Section \ref{sec:result} discusses overall properties of CDM and SIDM haloes (sub-section \ref{sec:overall}), the density profiles of DM and stars (\ref{sec:density}), the cosmological evolution of DM and stellar distribution (\ref{sec:history}), and the role of  baryon concentration to develop diverse DM central densities in SIDM (\ref{sec:diversity}). In section \ref{sec:summary} we summarize our results.

\section{Simulations}\label{sec:sim}
We perform a suite of cosmological simulations with virial masses $\sim 10^{12}\, \text{M}_\odot$ at $z=0$. Our simulations are part of the ``Feedback In Realistic Environments" project \citep[FIRE,][]{hopkins2014,hopkins2018}\footnote{\url{https://fire.northwestern.edu}}, and are run using {\sc GIZMO} \citep{hopkins2015}\footnote{\url{http://www.tapir.caltech.edu/~phopkins/Site/GIZMO.html}}. The gravity is solved with an improved version of the Tree-PM solver from {\sc GADGET3} \citep{springel2005} and the hydrodynamical equations are treated via the mesh-free Lagrangian-Godunov (MFM) method which provides adaptive spatial resolution while maintaining conservation of mass, energy, and momentum. 

The baryonic physics implementation in FIRE-2 includes cooling, star formation, stellar feedback including SNe Ia \& II, multi-wavelength photo-heating, stellar winds, radiation pressure, and UV-background radiation, all taken from stellar evolutionary models. Star formation happens in molecular gas clouds that are locally self-gravitating, self-shielding, Jeans unstable, and above the density threshold $n_{\rm H}>n_{\rm crit}=1000\, \text{cm}^{-3}$. A detailed description of the baryonic physics implementations can be found in \citet{hopkins2018}. Our simulations reach mass resolutions $m_{\rm b}=7.0\times 10^3\, \text{M}_\odot$ and $m_{\rm dm}=3.5\times 10^4\, \text{M}_\odot$ for the baryons and DM, and minimum physical spatial resolutions $\epsilon_{\rm gas}=1\, \text{pc}$, $\epsilon_\star=4\, \text{pc}$ and $\epsilon_{\rm dm}=20\, \text{pc}$.   

Initial conditions\footnote{\url{http://www.tapir.caltech.edu/~phopkins/publicICs/}} are generated using the zoom technique \citep{katz1993, onorbe2014} at $z=99$, embedded within periodic cosmological boxes of length $L=60\, {\rm Mpc}/h$ using the code {\sc MUSIC} \citep{hahn2011}. We adopt the following cosmological parameters, based on the final nine-year WMAP data \citep{wmap2013}: $\Omega_\Lambda=0.728$, $\Omega_{\rm m}=1-\Omega_\Lambda=0.272$, $\Omega_{\rm b}=0.0455$, $n_{\rm s}=0.963$, $\sigma_8=0.801$, and Hubble paramter $h=0.702$. We identify and quantify haloes and subhaloes in our simulations with a modified version of the code {\sc ROCKSTAR} \citep{behroozi2013,samuel2020}.

\begin{table*}
	\centering
	\caption{Galaxy properties. Columns describe: (1) Name: name and DM type of each simulation, (2) $M_{\rm 200}$: halo mass identified as the mass enclosed by the radius where cumulative density is $200$ times the critical density of the Universe ($\rho_{\rm crit}$), (3) $r_{200}$: the radius in which the mean density is equal to 200 times $\rho_{\rm  critical}$ (4) $V_{\rm max}$: maximum circular velocity, (5) $V_{\rm 2\, kpc}$: total circular velocity at $2\, {\rm kpc}$, (6) $\Phi_{\rm pot}$: total gravitational potential at the center of the halo (see section \ref{sec:diversity}), (7) $M_{\star}$: galaxy stellar mass, computed as the sum of stellar masses inside of $0.1*r_{200}$, (8) $r_{1/2}$: 3D stellar half-mass radius, (9) $\rho_{\rm DM}(0.5\, {\rm kpc})$: DM density profile at $500$ pc, in $\text{M}_\odot\, {\rm kpc}^{-3}$ (10) $\gamma(0.5\texttt{-}1.0\, {\rm kpc})$: logarithmic DM density slope measured between $0.5$ and $1$ kpc. $\dagger$: We stop this simulation at $z=0.1$ due to its extreme computational cost at low $z$. We have checked that the density and velocity dispersion profiles are very similar between $z=0.1$ to $z=0$ for other two m12 haloes.
	}
	\label{tab:properties}
	\setlength\tabcolsep{5pt}
	\begin{tabular*}{\textwidth}{lccccccccc}
	    Name & $M_{\rm 200}\, ({\rm M}_\odot)$ &$r_{200}\, ({\rm kpc})$ &$V_{\rm max}\, ({\rm km/s})$ & $V_{\rm 2\, kpc}$ & $\Phi_{\rm pot}\, ({\rm km}^2 {\rm s}^{-2})$ & $M_{\star}\, ({\rm M}_\odot)$ & $r_{1/2}\, ({\rm kpc})$ & $\rho_{\rm DM}\, (0.5{\rm kpc})$ & $\gamma(1\texttt{-}0.5{\rm kpc})$\\
	    \hline
		\hline
		m12i-CDM & $0.9\times10^{12}$ & 197 &187 & 160 & $-2.0\times10^5$&$3.4\times10^{10}$ & 4.3& $1.3\times10^8$ & -0.49\\
		m12i-CDM-only&$1.0\times10^{12}$ & 204&162 & 63 &$-2.2\times10^5$ &-  &  -  & $1.45\times10^8$& -0.94\\
		m12i-SIDM1 & $0.9\times10^{12}$ & 197 &214 & 207 &$-2.6\times10^5$ &$5.2\times10^{10}$ & 3.8  & $4.0\times10^8$ & -0.97\\
		m12i-SIDM10(z=0.1)$^\dagger$& $0.9\times 10^{12}$ & 190  &  211  & 207  &  $-2.7\times10^5$ &$4.7\times10^{10}$  &  3.6 &$8.0\times10^8$ & -1.3\\ 
		\hline
		m12f-CDM & $1.4\times 10^{12}$ & 228&222 & 221 &$-3.1\times10^5$ &$5.8\times10^{10}$ & 3.6 & $3.4\times10^8$ & -0.91\\
		m12f-CDM-only& $1.4\times10^{12}$&228 & 176 & 71 &$-2.6\times10^5$& -  &  -   & $2.0\times10^8$ & -1.09 \\
		m12f-SIDM1 & $1.4\times 10^{12}$ & 228&225 & 202 & $-2.8\times10^5$ &$6.7\times10^{10}$& 4.7 & $2.8\times10^8$ & -0.71\\
		m12f-SIDM10& $1.3\times10^{12}$ & 222&220 & 218 & $-2.6\times10^5$&$5.4\times10^{10}$& 3.5 & $4.6\times10^8$ & -0.92\\
		m12f-SIDM1-only& $1.4\times10^{12}$&228& 177 & 27 & $-2.3\times10^5$& -  &  -  & $1.0\times10^7$ & -0.03 \\
		m12f-SIDM10-only&$1.4\times10^{12}$&228&185&19& $-2.3\times10^5$& -  &  -   & $4.8\times10^6$ & 0.02\\
		\hline
		m12m-CDM &$1.2\times10^{12}$ & 217&199 & 118 & $-1.8\times10^5$&$5.0\times10^{10}$& 8.2 & $8.3\times10^7$ & -0.30\\
		m12m-CDM-only&$1.2\times10^{12}$&217&171&85& $-2.7\times10^5$& - & -  & $3.0\times10^8$ & -1.03\\
		m12m-SIDM1 &$1.2\times10^{12}$&217&221&126& $-2.0\times10^5$&$6.8\times10^{10}$ & 8.2& $8.0\times10^7$ & -0.16\\
		m12m-SIDM10&$1.2\times10^{12}$&217&226&204&$-2.7\times10^5$ &$8.2\times10^{10}$& 7.2 & $4.9\times10^8$ & -0.81\\
		\hline
	\end{tabular*}
\end{table*}

We take three m12 haloes (m12i, m12f, m12m) from the Latte suite \citep{wetzel2016,gk2019} and re-simulate them with the SIDM model with two constant DM cross sections $\sigma/m=1\,{\rm and}\, 10\, {\rm cm}^2\, {\rm g}^{-1}$ (labelled as SIDM1 and SIDM10). In our standard simulation suite any thermal-to-kinetic energy conversion from the stellar mass-loss processes are ignored for the sub-resolution regions \citep{hopkins2018b,hopkins2018}. As we discuss it further in section \ref{sec:overall}, this implementation could have significant effect on stellar masses and DM central densities (the same simulated galaxies with maximum stellar mass-loss energy conversion could have 2-3x higher stellar masses and denser DM density profiles). We also perform two DM-only (DMO) SIDM simulations with the aforementioned DM cross sections for the m12f halo for the purpose of comparison. The SIDM module in {\sc GIZMO} \citep[introduced in][]{rocha2013,peter2013} uses a Monte-Carlo approach to assign scattering probability for nearest neighbours of each DM particle via a spline kernel \citep{monaghan1985}, and then isotropically assigns velocities to the scattered particles such that energy and momentum are conserved.  

\section{Results}\label{sec:result}
\subsection{Mass Assembly and Star Formation Rate}\label{sec:overall}
It has been widely discussed in the literature that SIDM preserves the predictions of CDM on the scales larger than galaxies \citep[e.g. see][]{rocha2013,vog2016,sameie2019}. In this section, we discuss the impact of SIDM on the overall features of galaxies embedded in their DM haloes. Table \ref{tab:properties} summarizes halo and galaxy properties for the simulation suite; CDM and SIDM simulations have no notable difference in their virial masses, $M_{200}$\footnote{We define the virial mass to be the mass enclosed by the sphere with an average density of 200 times the critical density of the universe. The radius of this sphere is called virial radius, $r_{200}$.}, while SIDM simulations have higher stellar masses at $z=0$ compared to the CDM runs. In addition,  galaxy sizes, quantified via the 3D stellar half-mass radius $r_{1/2}$, are smaller in the CDM simulations compared to the SIDM1 runs, while the simulations with $\sigma/m=10$\sig show signs of contraction in their stellar distribution compared to the SIDM1 suite. The maximum total circular velocity, $V_{\rm max}$, is comparable or higher in the SIDM haloes than their CDM halo counterparts. A related quantity often used to characterize halo rotation curves, the total circular velocity computed at $2$ kpc ($V_{\rm 2\, kpc}$), is also similar or higher in the SIDM haloes than the CDM ones. Introducing baryonic processes in both CDM and SIDM simulations causes a systematic increase in both $V_{\rm max}$ and $V_{\rm 2\, kpc}$ compared to their DMO version. Interestingly, SIDM haloes respond more significantly to the presence of baryons with a proportionally greater increase in their $V_{\rm 2kpc,\, DMO}/V_{\rm 2kpc,\, Hydro. }$ of $\sim 0.10$ compared to $0.25\texttt{-}0.35$ for the CDM haloes.\\
\begin{figure*}
    \centering
    \includegraphics[width=\textwidth]{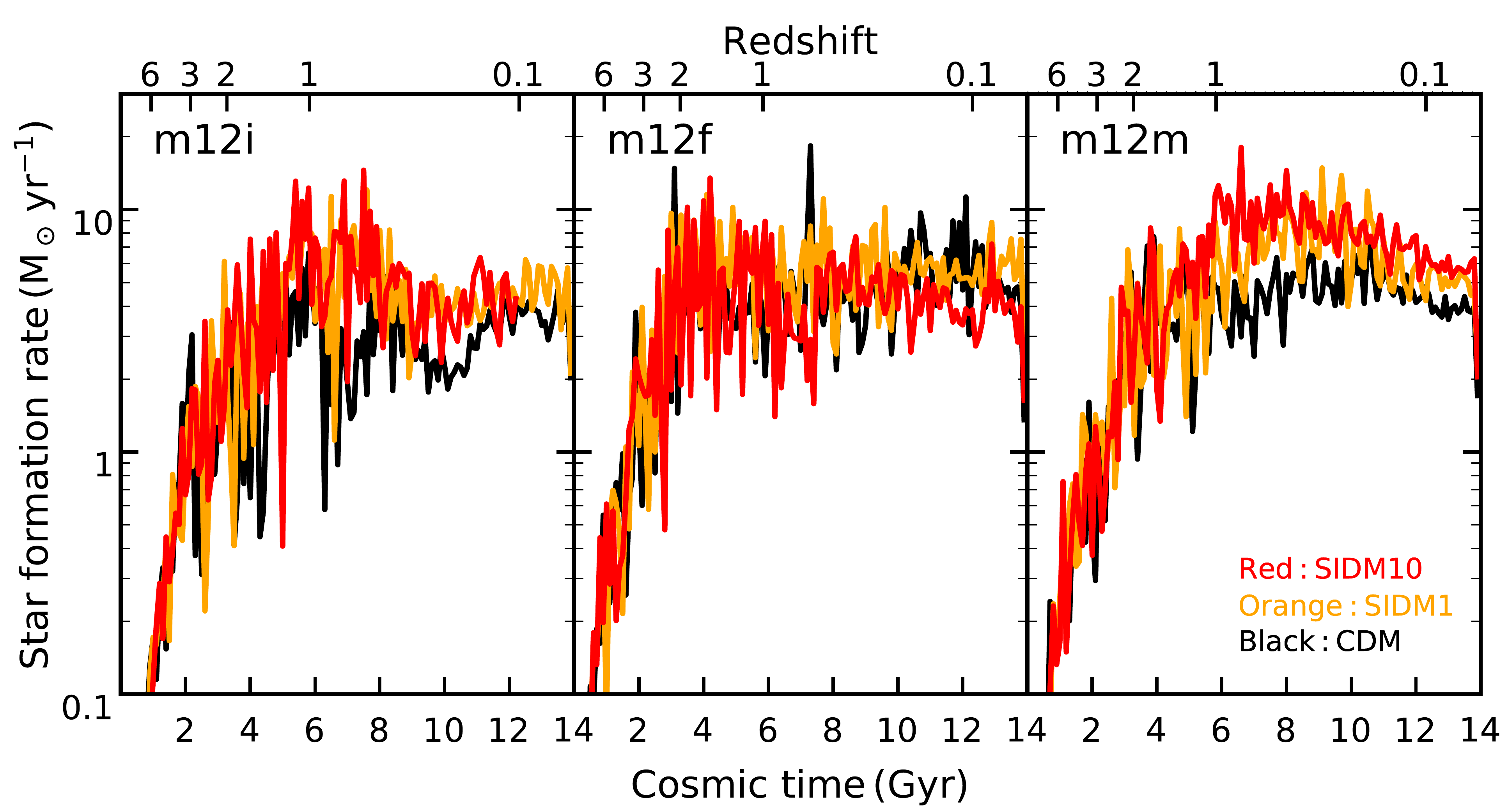}
    \caption{Star formation rate profiles for the CDM (black), SIDM1 (orange), and SIDM10 (red) models in m12i (left), m12f (middle), and m12m (right) galaxies. In order to compute the SFR profiles, we take all the star particles inside of $0.1*r_{200}$ of the host in the last snapshot, and we plot their star formation times in 100 Myr intervals.
    }
    \label{fig:sfr}
\end{figure*}
\indent We further explore the growth of stellar mass in CDM versus SIDM by comparing the star formation rates (SFR) of the simulated haloes in Fig. \ref{fig:sfr}. We compute the SFR by taking all star particles within $0.1*r_{200}$ of the host halo's center in the last snapshot, and plotting their formation times in 100-Myr intervals. All the simulated galaxies show bursty star formation with two distinct phases: an early phase when the SFR increases with time, followed by a constant and non-zero SFR that extends to the present time. However, after early, rapid phase of star formation, the SIDM models form stars at similar or higher rates compared to the CDM runs, leading to galaxies with higher stellar masses at $z=0$. The relative increase of the SFR in the SIDM models is different for each halo. SIDM runs in m12m haloes have a larger difference in their SFR relative to CDM, while the  SFR profiles of the m12i and m12f-SIDM haloes are closer to the corresponding CDM run.

\subsection{Density Profiles}\label{sec:density}
\begin{figure*}
    \centering
    \includegraphics[width=\textwidth]{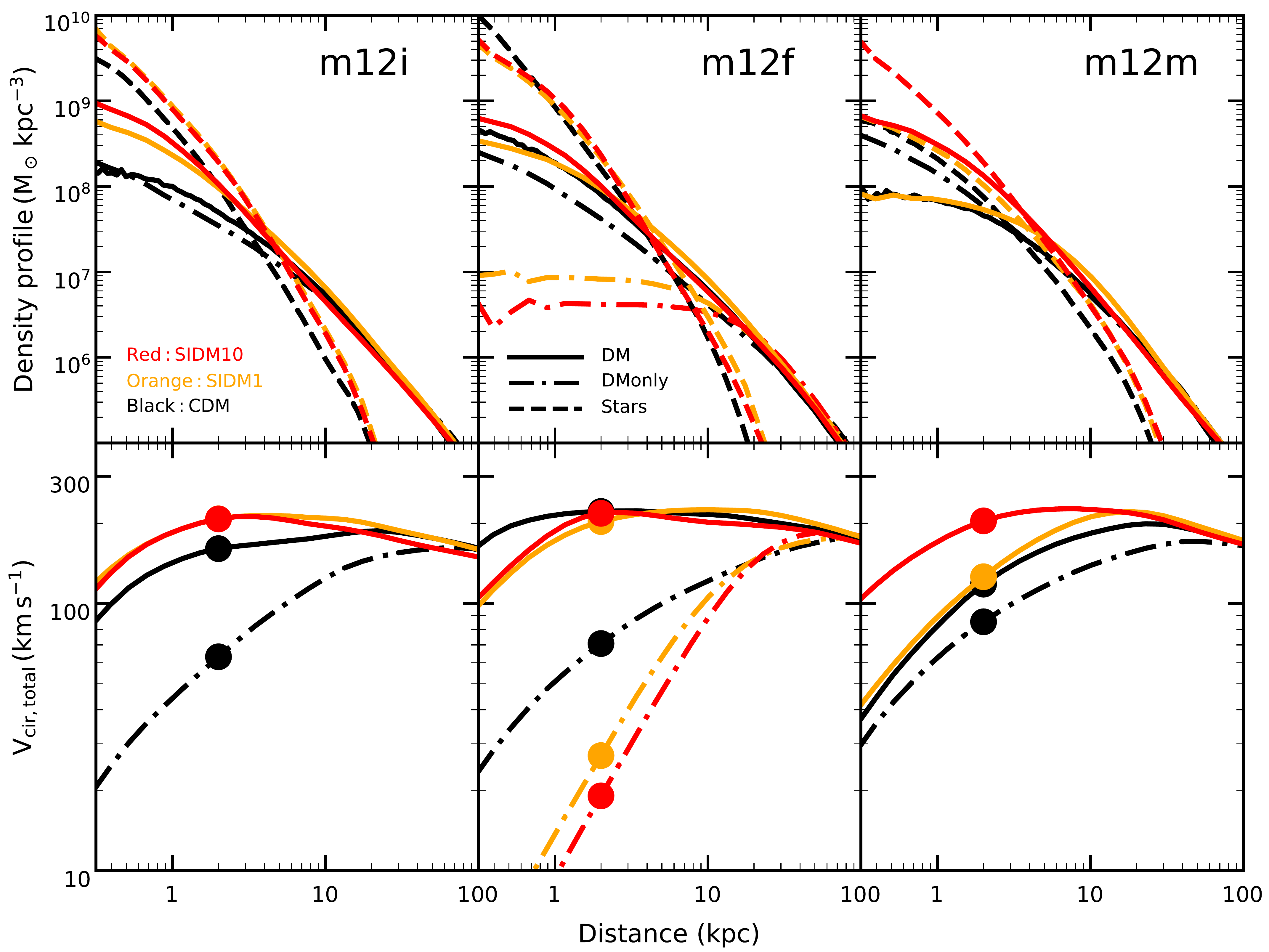}
    \caption{{\bf Top}: Spherically-averaged density profiles for DM (solid)and stellar  (dashed) particles for m12i (left), m12f (middle), and m12m (right) systems. The color code is the same as in Fig. \ref{fig:sfr}. We also plot available DMO runs for each halo (dot-dashed) for comparison. {\bf Bottom}: The corresponding total circular velocity profiles for the simulated systems in the top panel. The dots specify the circular velocity at $2$ kpc, i.e. $V_{\rm 2\, kpc}$. 
    }
    \label{fig:density}
\end{figure*}

Fig. \ref{fig:density} shows density profiles for DM (solid) and stars (dashed) for the simulated haloes in CDM (black), SIDM1 (orange), and SIDM10 (red). For comparison, DM density profiles of the available DMO runs are shown by dot-dashed lines.\footnote{DMO runs are identical to the hydro version except there are no baryons, so the baryonic mass is added to the DM particle mass. In order to have a direct comparison to DM density profiles from the hydro versions, we multiply the density by $1-\Omega_{\rm b}/\Omega_{\rm m}\sim0.83$ to account for the absence of the baryons.} The DMO versions of the CDM runs, predictably, show cuspy inner regions with the DM density slope $\gamma(r)\equiv{\rm dlog}\, \rho(r)/{ \rm dlog\, r}\sim -1.0$ at $r=0.5\texttt{-}1$ kpc, while the baryonic versions show flattening toward the center, with $-0.9\leq\gamma\leq-0.3$ (Table \ref{tab:properties}). SIDM models, on the other hand, have more diverse DM density slopes with $\gamma\sim 0$ for DMO runs and $-0.97\leq\gamma\leq-0.28$ for the full physics simulations (Table \ref{tab:properties}). 

The build-up of the stars in the baryonic versions of our simulations results in stellar components that dominate the gravitational potential in the inner regions. This causes all the haloes with baryons to develop central densities that are up to 100x higher than their DMO counterparts (except m12m with CDM), with SIDM haloes responding more dramatically compared to their CDM counterparts. The impact of stellar mass assembly is opposite for the CDM and SIDM models: in CDM haloes, stellar feedback flattens the density profile while in SIDM haloes central density becomes cuspier. Evidently, feedback does little to flatten the central density of the SIDM haloes; the combination of SIDM thermalization and baryonic contraction tends to make the DM density profiles less cored than in the DMO case. We note that the shape of DM density profiles is not universal among all SIDM haloes, and different realizations of MW-mass galaxies could have different central DM densities and slopes (Table \ref{tab:properties}). As we discuss in the next few sections, a combination of SIDM cross section, and star formation history and baryonic concentrations controls the evolution of the SIDM central densities.

Haloes with different SIDM cross sections show quite different responses to the stellar mass assembly. In all the SIDM haloes with $\sigma/m=10$\sig, the inner DM profile has higher central density and is cuspier than in the CDM and SIDM1 counterparts. This reflects the fact that higher DM cross section leads to higher interaction rate $\Gamma\propto \sigma/m$, and hence more efficient thermalization process. On the other hand, the three SIDM haloes with $\sigma/m=1$\sig show more comparable central densities to the CDM haloes: in m12i, the SIDM density profiles are both denser and cuspier than the CDM run, while the m12f-SIDM1 density profile is only slightly flatter compared to its CDM version, and the m12m run has an extended core with similar density. In all cases, higher (lower) central DM densities are accompanied by higher (lower) amplitudes for the stellar density profiles. The co-evolution of SIDM with $\sigma/m=1$\sig and baryons in our baryonic simulations generates slightly larger diversity in DM central density (measured by $\rho_{\rm DM}(0.5\, {\rm kpc})$; see table \ref{tab:properties}) compared to the CDM runs \citep[see also][]{elbert2018,sameie2018,robles2019}. This is very different from the DMO simulations, which result in constant central densities with amplitudes that decrease monotonically with cross section. 

In the lower row of Fig. \ref{fig:density}, total circular velocity profiles, $V_{\rm tot}=\sqrt{GM_{\rm tot}(<r)/r}=\sqrt{V_{\rm dm}^2+V_{\rm \star}^2+V_{\rm gas}^2}$, are plotted for the CDM and SIDM simulations. In order to reassess the diversity in the central DM distributions, we adopt the quantity $V_{\rm 2\, kpc}$, introduced by \citet{oman2015}, which is the total circular velocity at $2\, {\rm kpc}$ (filled circles). Again, SIDM models with $\sigma/m=1$\sig show a similar diversity in the distribution of $V_{\rm 2\, kpc}$ compared to the CDM haloes. The SIDM runs with cross section $10\, {\rm cm}^2\, {\rm g}^{-1}$ have less diversity in their $V_{\rm 2\, kpc}$ distribution, reflecting the very high central densities described above. The sample of our baryonic simulations suggest that SIDM models with cross section $\sigma/m=1$\sig, as well as the CDM+feedback, can generate diverse DM distributions in the MW-mass scale \citep[see also][]{santos2018}. 
A larger sample of galaxies simulated in both CDM and SIDM models, with more variety in their mass assembly and star formation history, and at different mass scales relevant to the diversity problem will be crucial to confirm our results. 

\subsection{Redshift evolution of density and velocity dispersion profiles}\label{sec:history}

\begin{figure*}
    \centering
    \includegraphics[width=\textwidth]{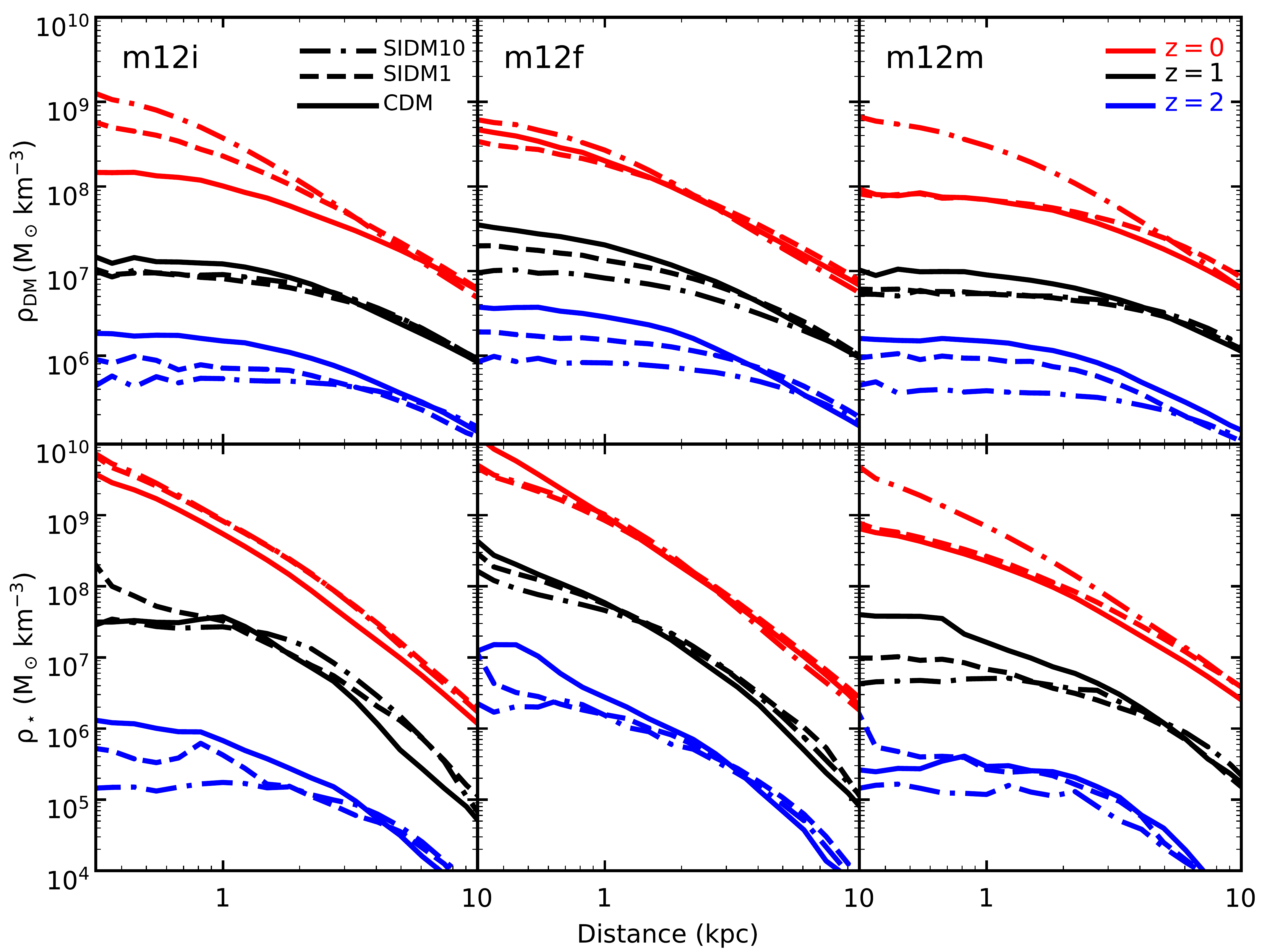}
    \includegraphics[width=\textwidth]{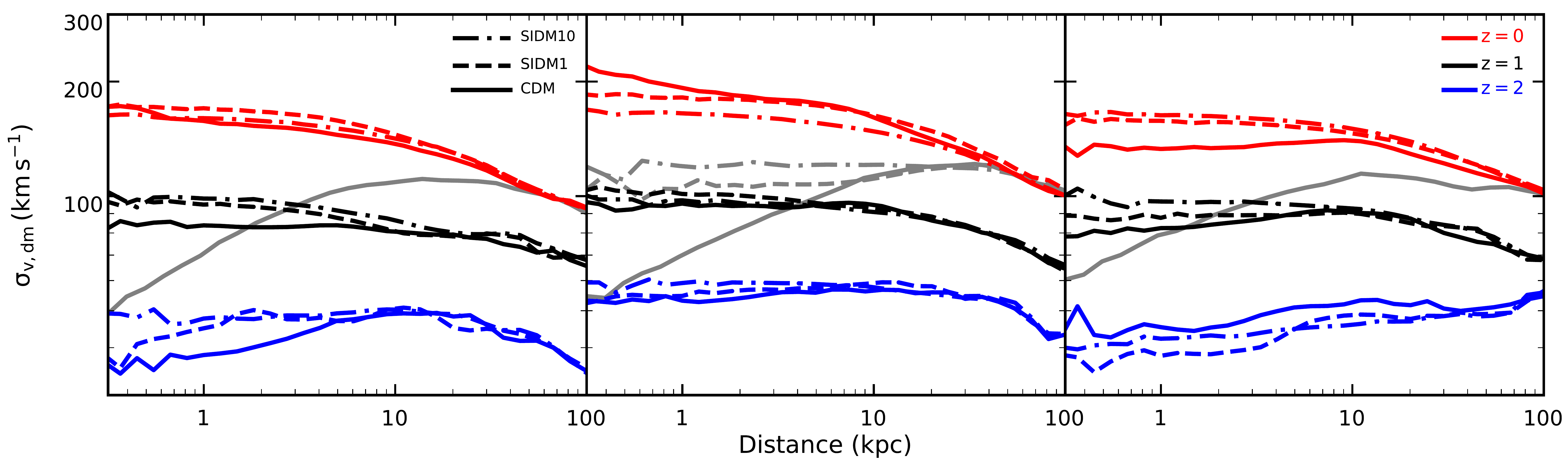}
    \caption{Cosmological evolution of DM density (top), stellar density (middle) and 1D DM velocity dispersion (bottom) for m12i (left column), m12f (middle column) and m12m (right column) haloes. Colored curves represent different redshifts. Density and dispersion profiles at $z>0$ are moved 1.5 and 0.75 dex downward with respect to each other for the purpose of clarity. In the bottom row, the DMO version of the simulated haloes are shown by grey curves at $z=0$.    }
    \label{fig:evolution}
\end{figure*}

In order to better understand the interplay between baryons and DM, Fig.~\ref{fig:evolution} shows the redshift evolution of the DM density (top), stellar density (middle), and DM velocity dispersion (bottom) profiles (all in physical coordinates) for the CDM and SIDM models.  Results for $z>0$ are shifted downward by $1.5$ and $0.75$ dex for density and dispersion profiles respectively, for clarity. At $z=2$ (blue curves), both CDM and SIDM models have already developed extended (several kpc) density cores, with core sizes increasing with the DM cross section. We note that at these redshifts our Milky Way-mass systems effectively are acting like dwarf galaxies in terms of core formation \citep[][, see section \ref{sec:diversity} for further discussion]{chan2015,elbadry2016,lazar2020}. The density profiles for SIDM1 and SIDM10 become more similar for $z\lesssim1$, and thereafter SIDM10 haloes develop higher central densities with smaller core sizes \citep[see also][]{sameie2018}. In fact, the higher DM self-interaction cross section in the SIDM10 models leads to more efficient heat transport (SIDM interaction rate $\Gamma\propto \sigma/m$) and a faster transition from core-expansion to core-contraction phase. 
 
A comparison of the three SIDM1 simulations reveals varying amounts of halo contraction in the central region. We identify different growth rates in the DM central densities to correlate with different stellar formation histories  of the inner regions. The middle row of Fig. \ref{fig:evolution} shows the redshift-evolution of the stellar density profiles. At $z=2$, the stellar distributions in the CDM simulations have vastly different central densities (nearly a factor 100) and inner slopes, reflecting the variety of star formation rates for each galaxy \citep{santistevan2020}. The stellar profiles in the SIDM model are less centrally dense and more diffuse at $z=2$ compared to the CDM galaxies, but show more diverse evolutionary phases at later stages. The most notable difference is in the m12i runs where the SIDM models with initially lower stellar densities at $z=2$, form more stars at later times; by $z=0$, their stellar profiles are denser near the center compared to the CDM run. The transition from cored to cuspy DM profiles happens faster by increasing SIDM cross section. All three SIDM10 models with lower DM central densities at $z=2$ end up with denser and cuspier density profiles than the CDM haloes by $z=0$, while two out of three SIDM1 models still have lower or similar central densities compared to the CDM models. The higher DM cross section also leads to higher stellar central densities in the SIDM10 galaxies (see section \ref{sec:diversity} for further discussion).
 
\begin{figure*}
    \centering
    \includegraphics[width=\textwidth]{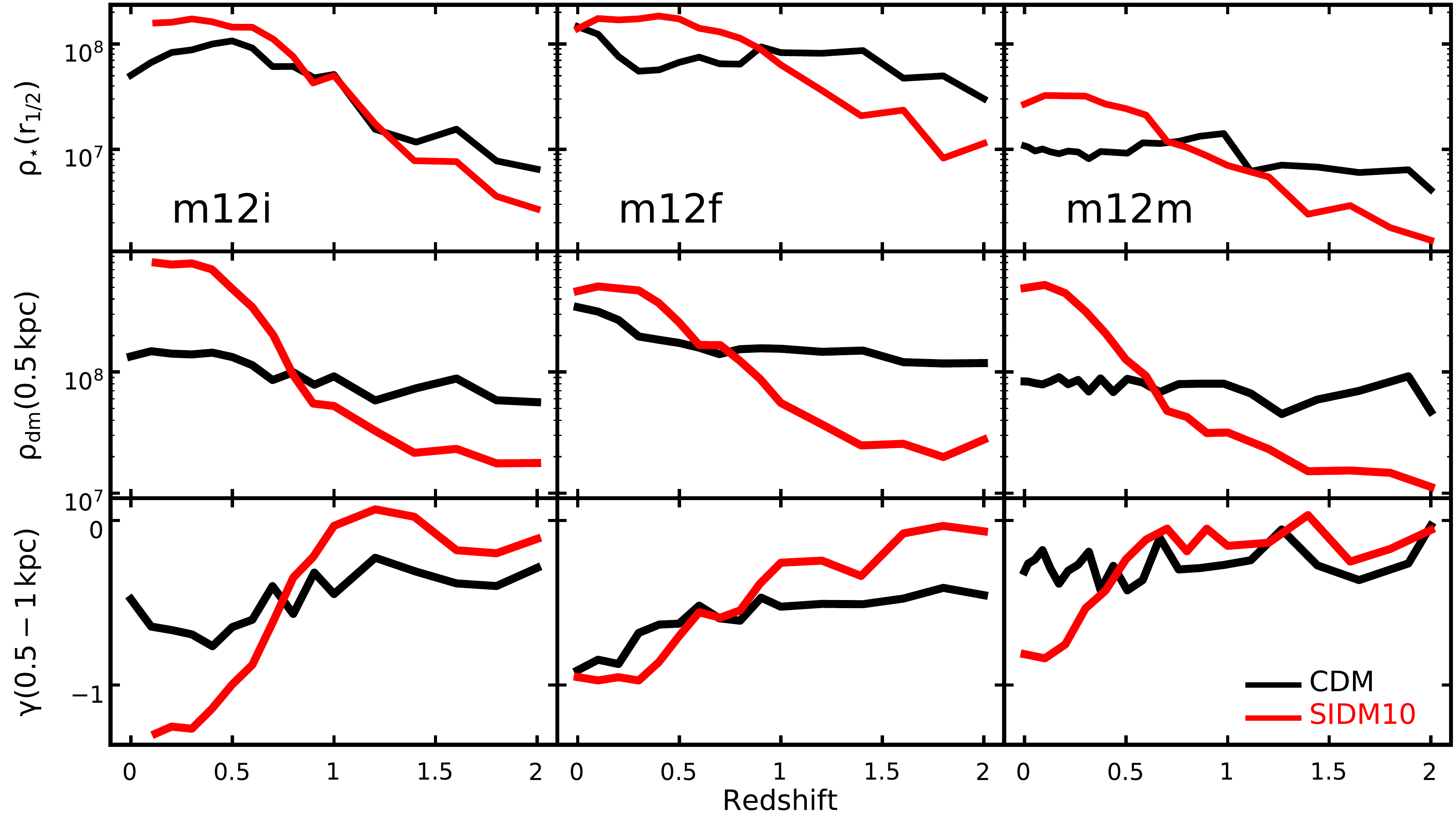}
    \caption{ {\bf Top}: The redshift evolution of the 3D stellar density, $\rho_\star$, computed at 3D half-mass radius $r_{1/2}$ for the CDM (black) and SIDM10 (red) galaxies since $z=2$.
    {\bf Middle}: DM density, computed at $r=0.5\pm0.05$ kpc for the CDM and SIDM10 haloes. {\bf Bottom}: DM density slope, computed between $r=0.5-1.0$ kpc for the CDM and SIDM10 haloes for all three realizations.}
    \label{fig:fig4}
\end{figure*}

The bottom row of Fig. \ref{fig:evolution} shows the evolution of the 1D DM velocity dispersion profiles for the CDM and SIDM models. For comparison, we also show $z=0$ DMO velocity dispersion profiles when available, as grey curves. There are two factors which contribute to the evolution of the velocity dispersion: the assembly of the DM and stars, and the thermalization of the halo due to DM self-interactions.  At $z=2$, both CDM and SIDM models have comparable velocity dispersion in all radii with positive slopes, indicating that DM self-interactions are yet to fully thermalize the central regions. Subsequent assembly of stars and DM increase the amplitude of velocity dispersion for both CDM and SIDM simulations. However, redistribution of energy and momentum in the SIDM haloes creates isothermal cores (i.e. flat central velocity dispersion). The relatively smaller halo-to-halo variation in the central velocity dispersion of the SIDM haloes signifies the role of thermalization process: while the central regions in the CDM haloes is heated up in response to the assembly of the baryons, the SIDM thermalization process redistributes heat such that the DM density cores remain isothermal. In addition, while the DM velocity dispersion in the baryonic versions of the SIDM simulations is higher than in their DMO versions (middle panel), the presence of baryons does not prevent the development of an isothermal core (i.e. flat velocity dispersion) induced by DM self-interactions\footnote{We have checked the detailed evolution of the DM velocity dispersion for both CDM and SIDM haloes since $z=1$. The dispersion profiles in both cases are almost identical, which in the case of SIDM implies the haloes have been isothermal for the last 7 Gyr.}. 

 We note that none of our SIDM haloes appear to be experiencing gravothermal core collapse. A necessary condition for SIDM core collapse is a negative gradient in the inner part of DM velocity dispersion profile such that the heat transport direction is inside-out. Our simulated SIDM haloes have velocity dispersion profile slopes consistent with ${\rm d}\log{\sigma_{\rm v}}/{\rm d}\log{r}=0$, which indicates that the cores are isothermal and are not collapsing. Moreover, the analytical model of \citet{essig2019} suggests the core-collapse time scale $t_{\rm c}$ scales as $t_{\rm c}\propto \big(\sigma/m\big)^{-1}M_{200}^{-1/3}c_{200} ^{-7/2}$. For a MW-mass halo with $\sigma/m=10\, {\rm cm}^2/{\rm g}$ and a halo concentration consistent with median of halo mass-concentration relation, the estimated core-collapse time is $t_{\rm c}=87$ Gyr, much longer than the $z=0$ Hubble time. High density peaks with extremely high halo concentration (>2$\sigma$ above the median $M_{200}\texttt{-}c_{200}$ relation) might experience the SIDM core collapse, but our small sample of simulated haloes with $\sigma/m=10\, {\rm cm}^2/{\rm g}$ does not contain any such haloes and is not large enough to study gravothermal core collapse. We will perform such an analysis in the future when we have completed a larger simulation suite with large SIDM cross sections.

In summary, we find that while in CDM the inclusion of baryons flattens the central density profile compared to the DMO version, in SIDM the dominance of baryons in the central regions can make the SIDM haloes cuspier than the CDM haloes. Either a denser stellar components or higher DM cross section lead to more cuspy SIDM profile, making the SIDM models more responsive to the presence of the baryons. Hence, the spread in the density of SIDM haloes is tied to the baryonic concentration and local star formation in the central few kpc of MW-mass galaxies. 

\subsection{The impact of baryon concentration on shaping DM density profiles}
\label{sec:diversity}

\begin{figure}
    \centering
    \includegraphics[width=\columnwidth]{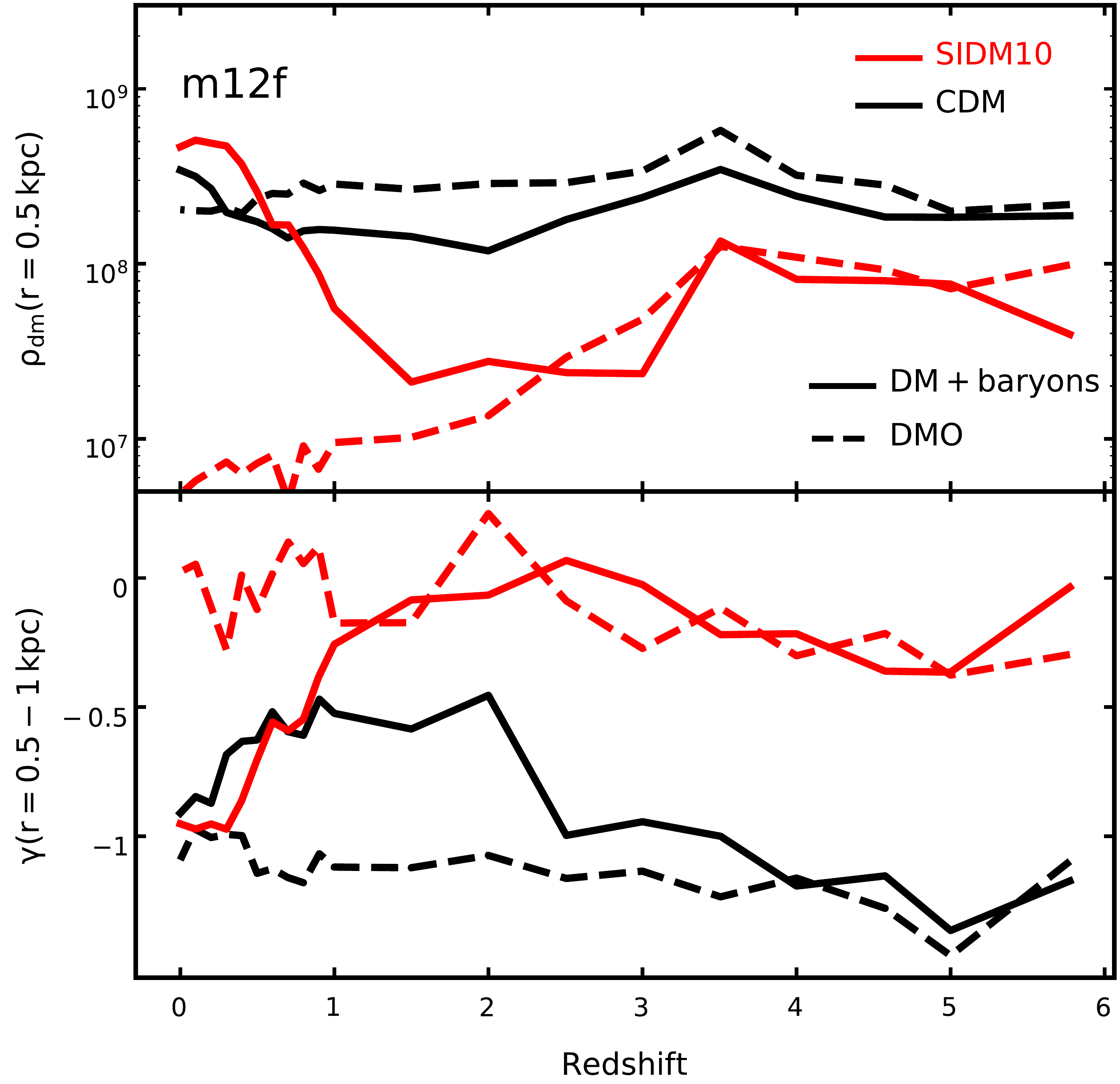}
    \caption{Redshift evolution of DM central densities computed at $r\texttt{=}0.5\pm0.05$ kpc (top) and the DM density slope measured at $r\texttt{=}1\texttt{-}0.5$ kpc (bottom), since $z\texttt{=}6$, for the m12f simulated galaxies with full baryonic physics (solid) DMO (dashed) in CDM (black) and SIDM10 (red). While the SIDM halos have cores even at the highest redshifts probed here, they end up with cuspy profiles at $z=0$ owing to the response of the SIDM halo to the stellar distribution at the center of the halo. Conversely, the CDM halo starts out cuspy (both with and without baryonic physics) and becomes cored after $z \approx 2$ owing to the effects of stellar feedback.}
    \label{fig:highz}
\end{figure}

Comparisons within our sample of simulated galaxies suggest that the concentration of the stellar distribution is more important than the total disc mass in creating diverse SIDM density profiles. Among SIDM1 simulations, the m12m galaxy is more massive than both m12i and m12f, and yet it has the most extended stellar disc with the largest stellar half-mass radius and lowest central DM density (table \ref{tab:properties}). The lower DM and baryonic concentration in the m12m SIDM1 halo gives rise to a central gravitational potential $\Phi_{\rm pot}(r\texttt{=}0)$ \footnote{We compute the gravitational potential by direct summation of all particles inside of virial radius, neglecting the contribution of boundary and distant particles.} that is shallower than the other two simulations (table \ref{tab:properties}). This picture is consistent with the analytical model of \citep{kaplinghat2014}, in which the SIDM density profile is related to the total gravitational potential by $\rho_{\rm SIDM}\propto \exp{(-\Phi_{\rm pot}/\sigma_{0}^2)}$ ($\sigma_0$ is the central DM velocity dispersion) and suggests that a larger (smaller) $\Phi_{\rm tot}$ leads to a higher (lower) central density in the SIDM haloes. In addition, controlled $N$-body simulations of SIDM haloes have shown that SIDM systems with higher baryonic concentration will transition faster from a ``core-expansion" phase to ``core-contraction" phase \citep{elbert2018,sameie2018}. 
 
As discussed in the previous section, SIDM10 models show such a transition from cored to cuspy profiles over the course of cosmological evolution (see top row panels of Fig. \ref{fig:evolution}). We find a correlation in the assembly of baryons and DM around the centers of our simulated galaxies. In the top panels of Fig. \ref{fig:fig4} we show the redshift evolution of the mean stellar densities $\rho_\star=M(<r)/(4\pi r^3/3)$ of the CDM (black) and SIDM10 (red) haloes computed at 3D half-mass radius $r_{1/2}$. The middle and bottom rows show the redshift evolution of DM central density computed at $r=0.5\pm 0.05$ kpc and the DM density slope computed at $r=0.5\texttt{-}1$ kpc \footnote{We have checked that our results are not sensitive to the radius we chose to compute DM densities, and the cumulative DM densities computed, for example, at 3D stellar half-mass radius also follow the same trend.}. At $z=2$, SIDM10 haloes have lower stellar densities and more cored and shallower DM density profiles than the CDM haloes. However, at lower redshift ($z\lesssim 1$), SIDM10 galaxies reach higher stellar densities, and their DM density profiles become denser and cuspier than the CDM simulations,  indicating that the SIDM haloes are in the core-contraction phase. 

It is interesting to observe that the stellar densities and DM density profile amplitudes in the SIDM10 haloes become higher than the CDM models around the same redshift, $z\sim 1.0\texttt{-}0.5$. The growth of the stellar densities in SIDM10 haloes come from both increase in the central stellar mass and decrease in $r_{1/2}$ relative to the CDM models; the latter effect is a response to SIDM halo contraction. A higher baryon concentration leads to higher SIDM interaction rates and more halo contraction. The decrease in the SIDM density amplitudes at $z<0.5$ is always accompanied by a decrease in the stellar central densities. This is consistent with analytical predictions where the SIDM density profiles are uniquely determined by the stellar gravitational potential in the regime $\Phi_\star\gg\Phi_{\rm dm}$ \citep{kaplinghat2014}. Our results are in agreement with the previous works and affirm that in the presence of massive and dense baryonic distribution the SIDM haloes can develop higher and more cuspy DM density profiles. In fully isolated halos, this combination of galaxy and halo contraction generates a run-away process that eventually leads to core collapse \citep{balberg2002,koda2011,sameie2018,essig2019}. 

Thus far, we have discussed the importance of baryon mass assembly in shaping the DM density profiles at late times ($z\lesssim2$). 
At higher redshifts, $z>2$, the baryonic potential does not contribute  significantly to the total gravitational potential, and therefore it cannot affect the evolution of central densities. In Fig. \ref{fig:highz} we compute the redshift evolution of DM central densities computed at $r\texttt{=}0.5\pm0.05$ kpc (top) and the DM density slope measured at $r\texttt{=}0.5\texttt{-}1$ kpc (bottom) for the m12f galaxies (solid) in CDM (black) and SIDM10 (red) models since $z=6$. We also compute these quantities for the DMO version of these haloes (dashed). At high redshifts, density profiles of the simulated haloes in full physics and DMO suites show good agreement in their amplitude and slope for both CDM and SIDM models. In the top panel, the central density in the CDM halo remains quite similar between full physics and DMO suite for the full redshift range plotted, while SIDM halo in the full physics suite shows a significant boost in its central density after $z=2$ (following a drop in density that comes from the effects of self-interactions). In the lower panel, the DM density slope of the full physics CDM system at $z=6$ (which has    $M_\star/M_{200}\sim 10^{-3}$ at that redshift) agrees well with its DMO version, while by $z=2$ ($M_\star/M_{200}\sim 10^{-2}$), the DM density profile becomes significantly flattened owing to stellar feedback. 

Our results for the CDM halo are consistent with those in \citet{lazar2020}, in which haloes with stellar-halo mass ratios of $M_\star/M_{200}\sim 10^{-3}$ have DM density slopes roughly equal to the DMO simulations, while haloes with $M_\star/M_{200}\sim 10^{-2}$ have much flatter DM density slopes. For the SIDM10 halo, the DM density slope also agrees well with its SIDM-only version at $z>3$; as a result of the self-interactions, both have much shallower inner density profiles compared to the CDM halo (see also Fig. \ref{fig:evolution}). At lower redshifts, the CDM and SIDM10 density profiles evolve strikingly differently: around $z=2$, the time at which the stellar feedback begins to flatten the DM density profile, the SIDM10 halo  starts to develop a steeper density profile (due to interplay between baryons and SIDM thermalization), and it eventually becomes cuspier than the CDM version of the same galaxy. Evidently, the core formation process is more controlled by the DM self-interactions than the stellar feedback in our SIDM simulation. Our findings are in line with previous works \citep{robertson2018, robles2019} which suggested once the DM halo becomes thermalized, the SIDM density profile is robust against potential fluctuations due to stellar feedback (e.g. see Fig. 1 of the first reference), or at the very least, it is not a dominant effect: the energy deposited by the stellar feedback is redistributed by frequent DM scattering such that the halo stays isothermal.

In summary, while the CDM central densities at $r=0.5$ kpc remain comparable between the full galaxy formation and DMO runs, stellar feedback leads to a shallower slope for density profile of the halo simulated with full baryonic physics at $z\lesssim2$ kpc. The central densities in the SIDM haloes become much larger and steeper owing to the interplay between baryons and DM self-interactions. Our results suggest that the core formation process in SIDM is controlled more by thermalization than baryonic physics, but  that the greater responsiveness of SIDM haloes (with $\sigma/m=10\,{\rm cm^2/g}$) to the presence of the baryons results in enhanced overall central densities relative to CDM.

\section{Conclusions}\label{sec:summary}
We perform a suite of baryonic cosmological zoom-in simulations with self-interacting dark matter (SIDM) models within the ``Feedback In Realistic Environment" (FIRE) project. The treatment of the baryonic physics includes cooling, star formation, stellar feedback from SNe, photo-heating, stellar winds, radiation pressure, and UV-background radiation. We take three Milky Way-mass dark matter (DM) haloes from the Latte suite and re-simulate them with SIDM models with constant DM cross sections $\sigma/m=1,\, 10$\sig. We study the stellar mass assembly and the evolution of DM density profiles in our SIDM simulations compared to the CDM versions. The CDM and SIDM haloes have almost identical halo assemblies with similar virial masses at $z=0$. Star formation rates (SFR) in both CDM and SIDM haloes are bursty and consist of two distinct phases: rapid increase in the SFR followed by a plateau from $z\leq1$ up to the present day. In the first phase of star formation, both CDM and SIDM simulations show similar SFRs. In the second phase, the SIDM haloes are more actively star-forming on average, and hence end up with more massive galaxies.

The interplay of star formation and the halo thermalization by DM self-interactions leaves distinct fingerprints on the slopes and central densities of DM distributions. Stellar feedback slightly flattens the slope of the DM densities in the CDM haloes compared to their DM-only counterparts, while SIDM density profiles become denser due to the assembly of the baryons. The build-up of stars in the CDM and SIDM1 haloes generates diverse DM central densities, while SIDM10 haloes show strong halo contraction due to more efficient SIDM thermalization process. 

A close scrutiny of the evolution of DM and stellar density profiles in the SIDM simulations reveals that once the stellar density profiles become significantly dense, the thermalization process becomes even more efficient, leading to faster core contraction in the SIDM haloes. Hence, the variation in star formation histories near the center of galaxies can lead to different DM central densities in the SIDM haloes. 

At high redshifts, $z>4$, CDM and SIDM haloes in both FIRE and DMO suite show good agreement in their central DM densities and slopes at high redshifts. However, at later times the CDM haloes tend to become more flattened (due to stellar feedback) while the our SIDM haloes become denser and cuspier. This different evolutionary phases in the CDM and SIDM suite mark the importance of the thermalization process by the DM self-interactions.

Overall, our results suggest that the SIDM haloes are creating diverse DM central densities from cored to cuspy density profiles and they could be even cuspier and denser than their CDM counterparts. We observe a correlation between the growth rate of central baryonic distribution and the DM central densities, and we find that SIDM haloes are more responsive to the presence of the baryons than the CDM haloes.

In order to examine the impact of specific treatments of stellar mass-loss on the larger-scale distribution of matter within the halos, we also perform CDM simulations for the m12i,f,m haloes that adopt a slightly different implementation of O/B and AGB-star photospheric outflows. In our default simulations, we ignore any conversion of thermal to kinetic energy (i.e., any ``$P\,{\rm d}V$ work'' done) during the Sedov-Taylor phase of the expansion from stellar mass-loss processes for the unresolved regions \citep[see][for more details]{hopkins2018b,hopkins2018}. Alternatively, stellar mass-loss processes in unresolved regions can be treated as a prolonged energy-conserving phase during which substantial $P\,{\rm d}V$ work was done, converting almost all of the thermalized/shocked ejecta energy into kinetic energy (momentum) on large scales.  This is the approach taken in standard FIRE-2 simulations and in the alternate stellar mass-loss simulations we discuss here: each stellar mass-loss event (which injects some $\Delta M \equiv \dot{M}_{\ast}\,\Delta t$, with initial free-streaming kinetic luminosity/energy $\Delta E \equiv \dot{E}\,\Delta t$) is considered as a ``mini-supernova'' and is treated in the exact same way as we do for SNe following \citet{hopkins2018b}.

The practical effect of adopting this (standard FIRE-2) treatment of stellar mass-loss, given the various scalings for, e.g., the cooling radii of SNe, is that most of total energy injection by stellar mass-loss is converted into momentum/kinetic energy on resolved scales (i.e.~${\sim}100\%$ of the stellar mass-loss energy is converted into macroscopic momentum), as compared to post-shock thermal energy, which can be more efficiently radiated away. This different implementation of the stellar mass-loss,  which is effectively stronger in terms of thermal-to-kinetic energy conversion in the sub-resolution region relative to our default simulations, serves as a useful comparison to understand the importance of the detailed treatment of stellar winds. It leads to higher SFR, more massive galaxies by a factor of 2-3, and less diverse DM density profiles with DM density slopes $\gamma=0.7-0.9$ and  DM central densities $\rho_{\rm DM}(0.5\,{\rm kpc})=(2.5-4.5)\times10^{8}\,{\rm M_{\odot}\,kpc^{-3}}$ and $V_{\rm 2\,kpc}\sim 250\,{\rm kpc}$ for all three haloes (see table \ref{tab:properties} for comparisons with our standard CDM runs) relative to the default model. Our results highlight the importance of sub-grid implementation of feedback models on creating diverse DM central densities, and we will present a more comprehensive analysis in a forthcoming paper.

\section*{Acknowledgements}
We thank Hai-Bo Yu for insightful discussions and the anonymous referee for a helpful and constructive report. MBK acknowledges support from NSF CAREER award AST-1752913, NSF grant AST-1910346, NASA grant NNX17AG29G, and HST-AR-15006, HST-AR-15809, HST-GO-15658, HST-GO-15901, HST-GO-15902, HST-AR-16159, and HST-GO-16226  from the Space Telescope Science Institute, which is operated by AURA, Inc., under NASA contract NAS5-26555.
RES acknowledges support from NASA grant 19-ATP19-0068, the Research Corporation for Scientific Advancement through Scialog-TDA, and grant HST-AR-15809 from the Space Telescope Science Institute, which is operated by AURA, Inc., under NASA contract NAS5-26555. Support for PFH was provided by NSF Research Grants 1911233 \&\ 20009234, NSF CAREER grant 1455342, NASA grants 80NSSC18K0562, HST-AR-15800.001-A. Numerical calculations were run on the Caltech compute cluster ``Wheeler,'' allocations FTA-Hopkins/AST20016 supported by the NSF and TACC, and NASA HEC SMD-16-7592. AW received support from NASA through ATP grants 80NSSC18K1097 and 80NSSC20K0513; HST grants GO-14734, AR-15057, AR-15809, and GO-15902 from STScI; the Heising-Simons Foundation; and a Hellman Fellowship. ASG is supported by the McDonald Observatory via the Harlan J. Smith postdoctoral fellowship. We ran simulations using: XSEDE, supported by NSF grant ACI-1548562; Blue Waters, supported by the NSF; Pleiades, via the NASA HEC program through the NASA Division at Ames Research Center. The analysis in this paper is carried out by python packages {\sc Numpy} \citep{vanderwalt2011}, {\sc matplotlib} \citep{hunter2007}, {\sc scipy} \citep{oliphant2007}, and {\sc h5py} \citep{collette2013}. 

\section*{Data availability}
The data supporting the plots within this article are available on reasonable request to the corresponding author.




\bibliographystyle{mnras}
\bibliography{bib} 


\appendix


\bsp	
\label{lastpage}
\end{document}